\documentclass[conference, 10pt]{IEEEtran}
\IEEEoverridecommandlockouts
\usepackage{cite}
\usepackage{amsmath,amssymb,amsfonts}
\usepackage{algorithmic}
\usepackage{graphicx}
\usepackage{textcomp}
\usepackage{xcolor}
\usepackage{subfigure}
\usepackage{stfloats}
\usepackage{amsfonts,subfigure,multicol,color,verbatim,
graphicx,cite,epsfig,amssymb,amsmath,cases,bm,algorithm,
algorithmic,xcolor,multirow,url,dsfont,bbm,tabularx}

\usepackage[left=0.63in, right=0.63in, top=0.75in, bottom=1in]{geometry}

\usepackage{setspace}

\newtheorem{lemm}{Lemma}

\newtheorem{prop}{Proposition}

\newcommand*{\QEDA}{\hfill\ensuremath{\blacksquare}} 

\IEEEoverridecommandlockouts
\IEEEpubid{\makebox[\columnwidth]{978-1-7281-4490-0/20/\$31.00 \copyright~2020~IEEE }
\hspace{\columnsep}\makebox[\columnwidth]{ }}

\begin{document}

\title{Intelligent Reflecting Surface Aided Multi-User mmWave Communications for Coverage Enhancement}

\author{\IEEEauthorblockN{Yashuai Cao\IEEEauthorrefmark{1}, Tiejun Lv\IEEEauthorrefmark{1} and Wei Ni\IEEEauthorrefmark{2}}
\IEEEauthorblockA{\IEEEauthorrefmark{1}Beijing University of Posts and Telecommunications, Beijing, China\\
\IEEEauthorrefmark{2}Data61, Commonwealth Scientific and Industrial Research, Sydney, NSW 2122, Australia\\
\{yashcao, lvtiejun\}@bupt.edu.cn, wei.ni@data61.csiro.au}
\thanks{
The financial support of the National Natural Science Foundation of China (NSFC) (Grant No. 61671072) and the Beijing Natural Science Foundation (No. L192025) is gratefully acknowledged.
}
}

\maketitle

\begin{abstract}
Intelligent reflecting surface (IRS) is envisioned as a promising solution for controlling radio propagation environments in future wireless systems.
In this paper, we propose a distributed intelligent reflecting surface (IRS) assisted multi-user millimeter wave (mmWave) system, where IRSs are exploited to enhance the mmWave signal coverage when direct links between base station and users are unavailable.
First, a joint active and passive beamforming problem is established for weighted sum-rate maximization. Then, an alternating iterative algorithm with closed-form expressions is proposed to tackle the challenging non-convex problem, thereby decoupling the active and passive beamforming variables. Moreover, we design a constraint relaxation technique to address the unit modulus constraints pertaining to the IRS. Numerical results demonstrate that the distributed IRS can potentially enhance the communication performance of existing wireless systems.
\end{abstract}


\section{Introduction}
The proliferation of applications and mobile traffic in fifth generation (5G) is driving the paradigm shift for wireless technologies, e.g., millimeter wave (mmWave) and massive multiple-input multiple-output (MIMO). However, providing the logical control of radio propagation remains as an unsettled challenge \cite{8811733}. For the envisioned mmWave systems with unrivaled data rates \cite{6847111}, signals are highly susceptible to blocking, thus causing sparse and low-rank channel structures.
To enhance the practical feasibility, such scheme as ultra-dense deployment \cite{7306533} also brings high cost and interference issues.

Recently, intelligent reflecting surface (IRS), as an emerging concept, has great potential to cost-effectively improve the network performance in 5G and beyond. Unlike the active amplify-and-forward (AF) relay, IRS is composed of nearly passive reflecting elements, and thus it is low cost, low power, flexible, and scalable \cite{8741198}. More importantly, the dynamic wireless environment can be turned into a partially deterministic space with the reconfiguration of the IRS.

The various IRS-aided wireless systems have been extensively studied. The reflect beamforming at the IRS is referred to as passive beamforming (PBF), while the precoding operation at the base station (BS) is termed as active beamforming (ABF). In \cite{8811733}, the ABF and PBF are jointly designed to reduce the transmit power while meeting the received signal-to-interference-plus-noise ratio (SINR) requirement for each user. Besides this, the future directions about the IRS deployment are also mentioned.
In \cite{8683145}, the beamforming design problems are extended to the discrete-phase cases.
Then, the authors of \cite{8723525} investigate the IRS-aided physical-layer security.
To study the suitability of IRS in terms of energy efficiency (EE), the authors of \cite{8741198} propose EE maximization algorithms for the multiple-input single-output (MISO) scenario. Most studies assume the rich scattering between BS and IRS \cite{8982186, 8746155}, but the low-rank BS-IRS channel \cite{8811733} should be considered when it comes to mmWave transmissions.
Recently, the potential use of IRS is studied in \cite{wang2019intelligent} from a perspective of mmWave, where weak BS-user links were compensated by the reflection gain of the IRS.

However, the low-rank BS-IRS channel cannot support multi-user downlink transmissions when the BS-user links are unavailable due to the severe blockage.
To conquer this issue, in this paper, we propose a distributed IRS (D-IRS) deployment solution to create higher-rank channels, and thus a smart radio environment created by multiple IRS units can achieve robust transmissions and reduce the signal blind spots.
In the proposed D-IRS enhanced multi-user mmWave systems without direct links, we formulate the weighted sum-rate maximization (WSM) problem by jointly designing the ABF and PBF matrices. This problem with the non-convexity and coupled variables is generally difficult. To tackle this issue, we propose an alternating iterative method, where closed-form expressions are derived in each iteration.
The non-convex unit-modulus constraints of the IRS constitute a great challenge of the PBF optimization. We thus devise a constraint relaxation approach to efficiently solve the PBF design. The effectiveness of our scheme is verified, and the impacts of network parameters on sum-rate performance are analyzed.
The main contributions of our work are as follows:
\begin{itemize}
\item For mmWave systems without direct links, a D-IRS deployment scheme is proposed to avoid low-rank BS-IRS mmWave channel. For one thing, this multi-IRS scenario breaks the constraint of rank-one channel, thus leading to high spatial multiplexing gain. For another, the smart radio environment created by multiple IRS units can achieve robust transmissions.
\item Based on the D-IRS enhanced multi-user mmWave system, we formulate the joint ABF and PBF design as a WSM problem. To tackle this non-convex problem, we propose an alternating iterative method, where closed-form expressions are derived in each iteration.
\item To tackle the unit modulus constraints, a constraint relaxation approach is devised. The performance of our scheme is evaluated, and the impact of the number of IRS elements as well as users on sum-rate performance is analyzed.
\end{itemize}

\emph{Notations}: Lower-case and upper-case boldface letters denote vector and matrix, respectively; $(\cdot)^{\ast}$, $(\cdot)^{\mathsf{T}}$, and $(\cdot)^{\mathsf{H}}$ represent the conjugate, the transpose, and the conjugate transpose; $\text{tr}(\cdot)$, $\text{vec}(\cdot)$ and $\text{diag}(\cdot)$ return the trace, vectorization and diagonalization; $[\cdot]_{i,j}$ represents the $(i,j)$-th entry of a matrix; $\jmath=\sqrt{-1}$; $\Re(\cdot)$ and $\arg(\cdot)$ denote the real part and the phase of a complex value; $\otimes$ and $\odot$ denote the Kronecker and Hadamard products, respectively.

\section{System Model}
\subsection{IRS-assisted Downlink mmWave MIMO System}
\begin{figure}[t]
	\centering{}\includegraphics[scale=0.38]{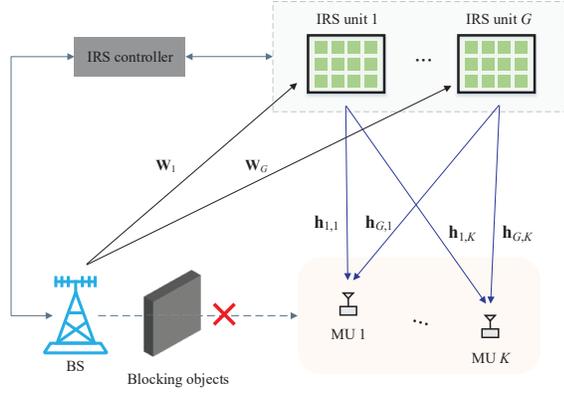}
	\caption{IRS-assisted downlink mmWave massive MIMO system.}
	\label{fig:system}
\end{figure}

Fig. \ref{fig:system} shows an IRS-assisted downlink mmWave system, where a base station (BS) is equipped with a uniform linear array (ULA) composed of $N$ elements. There are $G$ IRS units serving $K$ single-antenna mobile users (MUs). Each IRS unit is assumed to be with $M_{\rm az}$ elements horizontally and $M_{\rm el}$ elements vertically. Let $M=M_{\rm az} \times M_{\rm el}$. The direct BS-user links are assumed to be blocked by obstacles. The BS can only communicate with the MUs via the reflection of the IRS. These IRS units are controlled by a smart controller, which coordinates the reflecting modes for all IRS units.

The received signal at the $k$-th MU can be expressed as
\begin{align}
y_k =& \sum_{g=1}^{G}\mathbf h_{g,k}^{\sf H} \mathbf{\Phi}_g^{\sf H} \mathbf W_{g} \mathbf{p}_k {s}_k  \nonumber \\
&+ \sum_{g=1}^{G}\mathbf h_{g,k}^{\sf H} \mathbf{\Phi}_g^{\sf H} \mathbf W_{g} \sum_{j=1,j\neq k}^{K} \mathbf{p}_j {s}_j + {u}_k,
\label{eq:signal}
\end{align}
where $\mathbf W_{g} \in \mathbb C^{M \times N}$ is the mmWave channel between BS and the $g$-th IRS unit, and $\mathbf h_{g,k} \in \mathbb C^{M \times 1}$ is the channel between the $g$-th IRS unit and the $k$-th MU; the phase shift matrix of the $g$-th IRS unit is denoted by $\mathbf{\Phi}_g=\sqrt{\eta} \text{diag}([\theta_{g,1}, \cdots, \theta_{g,M}]^{\sf T})$ where $\eta$ is the reflection coefficient\footnote{Different from the backscatter communications, the incident signal energy is not absorbed to drive the circuit for IRS and thus we set $\eta=1$.} and $\theta_{g,m}=e^{\jmath \varphi_{g,m}}$ with $\varphi_{g,m}$ being the phase shift; $s_{k}$ is the transmit signal from BS for the $k$-th MU with zero mean and normalized power, and ${u}_k$ is the noise which follows the circularly symmetric complex Gaussian (CSCG) distribution of $\mathcal{CN}(0,\sigma_{\text u}^2)$; $\mathbf P = \left[ \mathbf p_1, \cdots, \mathbf p_{K} \right]$ is the precoding matrix\footnote{Undermined by mixed-signal processing and hardware constraints in practice, the hybrid precoding is a feasible scheme for the mmWave system. The subsequent hybrid precoding problem can be solved by the spatially sparse precoding \cite{6717211}. However, this is out of the scope of current manuscript.} where $\mathbf p_k = \in \mathbb C^{N \times 1}$ is used by BS to transmit signal $s_{k}$.
The total transmit power constraint at the BS is $\text{tr}(\mathbf P \mathbf P^{\sf H})\leq P_{\max}$.

\subsection{Wireless Channel Model}
According to the widely used 3D Saleh-Valenzuela channel model \cite{6834753}, $\mathbf W_{g}$ can be expressed as
\begin{align}
\mathbf W_{g} = \sum_{\ell=0}^{N_{\rm p}} \nu^{(\ell)} \mathbf{a}_{\rm{B}}\left(\phi^{(\ell)}_{\rm{B}}\right) \mathbf{a}_{\rm{I}}^{\sf{H}}\left(\phi^{(\ell)}_{\rm{I}}, \theta^{(\ell)}_{\rm{I}}\right), \label{eq:W_g}
\end{align}
where $N_{\rm p}$ denotes the number of non-line-of-sight (NLoS) paths, $\ell=0$ represents the line-of-sight (LoS) path, and $\nu^{(\ell)}$ is the complex gain of the $\ell$-th path. Note that mmWave channels normally consist of only a small number of dominant multipath components \cite{6847111, 7306533}, while the scattering at sub-6 GHz is generally rich. Here, the elevation and azimuth angles for two-dimensional IRS are denoted by $\theta^{(\ell)}$ and $\phi^{(\ell)}$. In (\ref{eq:W_g}), $\mathbf{a}_{\rm{B}}\left( \phi  \right) = \frac{1}{\sqrt{N}} \left[e^{-\jmath \frac{2\pi d}{\lambda} \phi i } \right]_{i \in \mathcal I(N)}$ and $\mathbf{a}_{\rm{I}}\left(\phi, \theta \right) = \mathbf{a}_{\rm{I}}^{\rm{az}}\left(\phi \right) \otimes \mathbf{a}_{\rm{I}}^{\rm{el}}\left(\theta \right)$ are the array steering vectors of ULA and IRS. The array steering vectors $\mathbf{a}_{\rm{I}}^{\rm{az}}\left(\phi \right)$ and $\mathbf{a}_{\rm{I}}^{\rm{el}}\left(\theta \right)$ are defined in the same manner as $\mathbf{a}_{\rm{B}}(\phi)$, where $\lambda$ is the wavelength, $d$ is the antenna spacing, and $\mathcal I(N)=\{n-(N-1)/2, n=0,1,\cdots,N-1\}$. The array element spacing of both ULA and IRS is assumed to be $\lambda/2$, and IRS is implemented with discrete antenna elements \cite{8580675, 8741198}, just like uniform rectangular array (URA).

Typically, the IRS is densely distributed in the hotspot spaces, which gives rise to a high probability of LoS path. Due to the severe path loss, the transmit power of 2 or more reflections can be ignored so that only LoS is considered \cite{8811733, 8683145}.
Thus, the channel between the $g$-th IRS unit and the $k$-th MU is given by
\begin{equation}
\mathbf h_{g,k}= \sqrt{M} \nu_{k} \varrho_{\text{r}} \varrho_{\text{t}} \mathbf a_{\text{t}}\left(\phi_{k}\right),
\label{eq:IRS_user}
\end{equation}
where $\nu_k$ is the channel gain; $\varrho_{\text{r}}$ and $\varrho_{\text{t}}$ are the receive and transmit antenna element gains, respectively; $\mathbf a_{\text{t}}$ is the array steering vector of IRS.
As typically considered in \cite{8811733, wang2019intelligent, 8741198}, all channels involved are assumed to be perfectly known in the paper.

\section{Joint Active and Passive Beamforming Design}\label{section:design}
\subsection{Problem Formulation and Transformation}
We aim to maximize the downlink sum-rate of the proposed D-IRS assisted mmWave network by jointly optimizing the ABF and PBF matrices.
The signal-to-interference-plus-noise ratio (SINR) of the $k$-th MU is given by
\begin{equation}
\gamma_k = \frac{\left\vert \sum_{g=1}^{G}\mathbf h_{g,k}^{\sf H} \mathbf{\Phi}_g^{\sf H} \mathbf W_{g} \mathbf{p}_k  \right\vert^2}{\sum_{j=1,j\neq k}^{K} \left\vert \sum_{g=1}^{G}\mathbf h_{g,k}^{\sf H} \mathbf{\Phi}_g^{\sf H} \mathbf W_{g}  \mathbf{p}_j \right\vert^2 + \sigma_{\text u}^2}.
\label{eq:sinr}
\end{equation}
The corresponding WSM problem can be formulated as
\begin{subequations}
\label{eq:fopt}
\begin{align}
\underset{\mathbf{P}, \mathbf{\Phi}_g}{\max} \quad & f_1(\mathbf{P}, \mathbf{\Phi}_g) = \sum_{k=1}^{K} \omega_{k} \log_2(1+\gamma_k)  \label{eq:opt1}
\\
\text{s.t.} \quad &\text{tr}\left(\mathbf{P} \mathbf{P}^{\sf H}\right) \leq P_{\max},  \label{eq:MIN1}\\
&  \theta_{g,m} \in \mathcal F_{\text{c}}, \quad \forall g,\ \forall m, \label{eq:MIN2}
\end{align}
\end{subequations}
where $\omega_{k}$ is the weight assigned to the data of the $k$-th MU, $P_{\max}$ is the maximum transmit power, and the continuous-phase feasible set for $\theta_{g,m}$ is $\mathcal F_{\text{c}}=\{\theta_{g,m} = e^{\jmath \varphi_{g,m}} \vert \varphi_{g,m} \in [0, 2\pi)\}$. Its discrete-phase counterpart is $\mathcal F_{\text{d}} = \left\{ \theta_{g,m} = e^{\jmath \varphi_{g,m}} \Big\vert\varphi_{g,m} \in \Big\{  \frac{2\pi i}{2^B}  \Big\}_{i=0}^{2^B-1} \right\}$, where $B$ is the phase resolution in the number of bits.

\begin{prop}
The WSM problem (\ref{eq:opt1}) is equivalent to
\begin{align}
\underset{\mathbf{P}, \mathbf{\Phi}_g, \bm \alpha}{\max} \ f_2(\mathbf{P}, \mathbf{\Phi}_g, \bm \alpha) =& \frac{1}{\ln 2} \sum_{k=1}^{K} \omega_{k} \ln(1+\alpha_k) \nonumber\\
&- \omega_{k} \alpha_k +  \frac{\omega_{k}(1+\alpha_k) \gamma_k}{1+\gamma_k},
\label{eq:sumR2}
\end{align}
where $\bm \alpha = \left[\alpha_1, \cdots, \alpha_K \right]^{\sf T}$ is the auxiliary vector introduced by the Lagrangian dual transform \cite{8314727}.
\end{prop}

\begin{IEEEproof}
See Appendix \ref{prf1}.
\end{IEEEproof}

Thus, solving (\ref{eq:opt1}) is equivalent to solving (\ref{eq:sumR2}).
We propose to solve (\ref{eq:sumR2}) through an alternating iterative approach, which results in three alternating steps, i.e., the auxiliary variable $\alpha_k$, ABF and PBF matrix set $\{\mathbf{\Phi}_1,\cdots, \mathbf{\Phi}_G\}$.

Setting $\frac{\partial f_2}{\partial \alpha_k}$ to zero yields $\hat{\alpha}_k=\gamma_k$.
Thus, when given $\mathbf{P}$ and $\mathbf{\Phi}_g$, $\alpha_k$ can be updated by solving (\ref{eq:sinr}) in each iteration.
Given $\alpha_k$, optimizing $\mathbf{P}$ and $\mathbf{\Phi}_g$ in (\ref{eq:sumR2}) can be recast to
\begin{align}
\underset{\mathbf{P}, \mathbf{\Phi}_g}{\max} \quad & f_3(\mathbf{P}, \mathbf{\Phi}_g)=\sum_{k=1}^{K} \frac{\omega_{k}(1+\alpha_k) \gamma_k}{1+\gamma_k}  \label{eq:opt2}
\\
\text{s.t.} \quad & \text{(\ref{eq:MIN1}), (\ref{eq:MIN2})}. \nonumber
\end{align}
By transforming to (\ref{eq:opt2}), the logarithm in (\ref{eq:opt1}) can be dealt with tactfully. In the following parts, we can alternately optimize $\mathbf P$ and $\{\mathbf{\Phi}_1,\cdots, \mathbf{\Phi}_G\}$ by solving (\ref{eq:opt2}).

\subsection{Active Beamforming Scheme}
We start with the ABF when given the PBF matrix set $\{\mathbf{\Phi}_1,\cdots, \mathbf{\Phi}_G\}$. For notational brevity, we define
\begin{equation}
\tilde{\mathbf h}_k^{\sf H} = \sum_{g=1}^{G}\mathbf h_{g,k}^{\sf H} \mathbf{\Phi}_g^{\sf H} \mathbf W_{g}.
\label{eq:hk}
\end{equation}
Substituting (\ref{eq:hk}) into (\ref{eq:sinr}), we reformulate $f_3$ as
\begin{align}
\underset{\mathbf{P}}{\max} \quad & f_4(\mathbf{P}) = \sum_{k=1}^{K} \frac{\bar{\alpha}_k \vert  \tilde{\mathbf h}_k^{\sf H} \mathbf{p}_k \vert^2}{\sum_{j=1}^{K} \vert \tilde{\mathbf h}_k^{\sf H}  \mathbf{p}_j \vert^2 + \sigma_{\text u}^2}
\label{eq:opt3} \\
\text{s.t.} \quad & \text{(\ref{eq:MIN1})}, \nonumber
\end{align}
where $\bar{\alpha}_k = \omega_{k}(1+\alpha_k)$.
We note that (\ref{eq:opt3}) is a multi-ratio fractional programming problem. By using the quadratic transform (QT) technique \cite{8314727}, we reformulate $f_4(\mathbf{P})$ as
\begin{align}
f_5(\mathbf{P}, \bm \beta) =& \sum_{k=1}^{K}  2\sqrt{\bar{\alpha}_k} \Re\{ \beta_k^{\ast}\tilde{\mathbf h}_k^{\sf H} \mathbf{p}_k \}
\nonumber \\
&- \vert\beta_k\vert^2 \bigg(\sum_{j=1}^{K} \vert \tilde{\mathbf h}_k^{\sf H}  \mathbf{p}_j \vert^2 + \sigma_{\text u}^2 \bigg)  ,
\label{eq:opt4}
\end{align}
where $\bm \beta = [\beta_1, \cdots, \beta_K]^{\sf T}$ is the auxiliary vector introduced by the use of QT, and (\ref{eq:opt4}) is a biconvex optimization problem.
By setting $\partial f_5 / \partial \beta_k=0$, the optimal $\beta_k$ for a given $\bf P$ is given by
\begin{align}
\hat{\beta}_k =  \frac{\sqrt{\bar{\alpha}_k} \tilde{\mathbf h}_k^{\sf H} \mathbf{p}_k}{\sum_{j=1}^{K} \vert \tilde{\mathbf h}_k^{\sf H}  \mathbf{p}_j \vert^2 + \sigma_{\text u}^2}.
\label{eq:optimal_beta}
\end{align}

Since Problem (\ref{eq:opt4}) is a convex problem with respect to $\mathbf p_k$, the optimal $\mathbf p_k$ can be obtained by the Lagrangian multiplier method. Thus, the optimal $\mathbf p_k$ for a given $\bm \beta$ is
\begin{align}
\hat{\mathbf p}_k = \sqrt{\bar{\alpha}_k} {\beta}_k \bigg(\mu \mathbf I_N + \sum_{i=1}^{K} \vert \beta_i \vert^2 \tilde{\mathbf h}_i \tilde{\mathbf h}_i^{\sf H} \bigg)^{-1} \tilde{\mathbf h}_k.
\label{eq:optimal_p}
\end{align}
where $\mu \ge 0$ is the Lagrange multiplier for the power constraint (\ref{eq:MIN1}). According to the following lemma, we can obtain the optimal $\mu$ via the bisection search method.

\begin{lemm}
The optimal value of $\mu$ in (\ref{eq:optimal_p}) is
\begin{equation}
\hat{\mu}=
\left\{
\begin{matrix}
{0} & {\begin{split}
\text{if}\ \textup{tr}(\mathbf{P} \mathbf{P}^{\sf H}) \le P_{\max} \\ \text{when} \ \mu=0
\end{split}} \\
\{\mu > 0: \textup{tr}(\mathbf{P} \mathbf{P}^{\sf H}) = P_{\max} \} & {\begin{split}
\text{if} \ \textup{tr}(\mathbf{P} \mathbf{P}^{\sf H}) > P_{\max} \\ \text{when} \  \mu=0
\end{split}}
\end{matrix}
\right.
\label{eq:optimal_mu}
\end{equation}
\end{lemm}

\begin{IEEEproof}
See Appendix \ref{prf2}.
\end{IEEEproof}

\subsection{Passive Beamforming Scheme}\label{section:pbf}
After the completion of ABF, we continue to solve the PBF problem pertaining to the IRS. With mathematical manipulations, (\ref{eq:hk}) can be rewritten as
\begin{align}
\tilde{\mathbf h}_k^{\sf H} \mathbf{p}_j &= \sqrt{\eta} \sum_{g=1}^{G} \bm \theta_g^{\sf H} \text{diag}\left( \mathbf h_{g,k}^{\sf H} \right) \mathbf W_{g} \mathbf p_j,
\label{eq:mani}
\end{align}
where $\bm \theta_g = [\theta_{g,1}, \cdots, \theta_{g,M}]^{\sf T}$. For notational brevity, we define $\mathbf v_{g,k,j} = \sqrt{\eta} \text{diag} ( \mathbf h_{g,k}^{\sf H} ) \mathbf W_{g} \mathbf p_j$.
Given $\bm \alpha$ and $\bf P$, we rewrite Problem (\ref{eq:opt3}) as
\begin{subequations}
\label{eq:opt5}
\begin{align}
\underset{\bm{\theta}_g}{\max} \quad & f_6(\bm{\theta}_g) = \sum_{k=1}^{K} \frac{\bar{\alpha}_k \left\vert \sum_{g=1}^{G} \bm \theta_g^{\sf H}  \mathbf v_{g,k,k} \right\vert^2}{\sum_{j=1}^{K} \left\vert \sum_{g=1}^{G}\bm \theta_g^{\sf H}  \mathbf v_{g,k,j} \right\vert^2 + \sigma_{\text u}^2} \\
\text{s.t.} \quad & \vert\theta_{g,m}\vert^2 = 1, \quad \forall g,\ \forall m. \label{eq:unit_mod}
\end{align}
\end{subequations}
To facilitate the subsequent derivations, we first construct
\begin{align}
\mathbf \Theta &= \left[\bm \theta_1, \bm \theta_2, \cdots, \bm \theta_G \right],  \label{eq:v21} \\
\mathbf V_{k,j} &= \left[\mathbf v_{1,k,j}, \mathbf v_{2,k,j}, \cdots, \mathbf v_{G,k,j} \right]. \label{eq:v22}
\end{align}
Then, $f_6(\bm{\theta}_g)$ can be reformulated as
\begin{align}
\underset{\underline{\bm\theta}}{\max} \quad  f_7(\underline{\bm\theta}) &= \sum_{k=1}^{K} \frac{\bar{\alpha}_k \left\vert \text{tr} \left(\mathbf \Theta^{\sf H} \mathbf V_{k,k} \right)\right\vert^2}{\sum_{j=1}^{K} \left\vert \text{tr} \left(\mathbf \Theta^{\sf H} \mathbf V_{k,j}\right) \right\vert^2 + \sigma_{\text u}^2}  \nonumber \\
&= \sum_{k=1}^{K} \frac{\bar{\alpha}_k \vert {\underline{\bm\theta}}^{\sf H} \underline{\mathbf v}_{k,k} \vert^2}{\sum_{j=1}^{K} \vert {\underline{\bm\theta}}^{\sf H} \underline{\mathbf v}_{k,j} \vert^2 + \sigma_{\text u}^2}.
\label{eq:opt6}
\end{align}
where $\underline{\bm\theta} = \text{vec} (\mathbf \Theta)$ and $\underline{\mathbf v}_{k,j}=\text{vec}(\mathbf V_{k,j})$. The corresponding QT of (\ref{eq:opt6}) is given by
\begin{align}
\underset{\underline{\bm\theta}, \bm \rho}{\max} \quad  f_8(\underline{\bm\theta}, \bm\rho) &= \sum_{k=1}^{K} 2\sqrt{\bar{\alpha}_k} \Re\left\{ \rho_k^{\ast} \underline{\bm\theta}^{\sf H} \underline{\mathbf v}_{k,k} \right\} \nonumber \\
&- \vert \rho_k \vert^2 \bigg(\sum_{j=1}^{K} \vert \underline{\bm\theta}^{\sf H} \underline{\mathbf v}_{k,j} \vert^2 + \sigma_{\text u}^2 \bigg) ,
\label{eq:opt7}
\end{align}
where $\bm \rho=[\rho_1, \cdots, \rho_K]^{\sf T}$ is the auxiliary vector introduced by the QT. Based on the Lagrange multiplier method, the optimal $\rho_k$ is given by
\begin{equation}
\hat{\rho}_k = \frac{\sqrt{\bar{\alpha}_k} \underline{\bm\theta}^{\sf H} \underline{\mathbf v}_{k,k}  }
{\sum_{j=1}^{K}  \vert \underline{\bm\theta}^{\sf H} \underline{\mathbf v}_{k,j} \vert^2 + \sigma_{\text u}^2}.
\label{eq:rho}
\end{equation}
Given $\bm \rho$, the optimization of $\underline{\bm\theta}$ can be written as
\begin{align}
\underset{\underline{\bm\theta}}{\max} \quad  f_9(\underline{\bm\theta}) &= -\underline{\bm\theta}^{\sf H} \mathbf A \underline{\bm\theta} + 2\Re\{\underline{\bm\theta}^{\sf H} \mathbf b\} - \sum_{k=1}^{K} \vert \rho_k \vert^2 \sigma_{\text u}^2,
\label{eq:opt8}
\end{align}
where $\mathbf A = \sum_{k=1}^{K} ( \vert \rho_k \vert^2 \sum_{j=1}^{K} \underline{\mathbf v}_{k,j}\underline{\mathbf v}_{k,j}^{\sf H} )$ and $\mathbf b = \sum_{k=1}^{K} \sqrt{\bar{\alpha}_k} \rho_k^{\ast} \underline{\mathbf v}_{k,k}$.
We can see that (\ref{eq:opt8}) is a quadratical constraint quadratic programming (QCQP) problem.

The IRS phase shift constraint (\ref{eq:MIN2}) is non-convex. We propose a constraint relaxation technique to address this unit-modulus constraint. Specifically, we replace (\ref{eq:unit_mod}) with $\vert\theta_{g,m}\vert^2 \le 1$ to convexity the constraint. By dropping the constant terms, we can rewrite (\ref{eq:opt8}) as
\begin{subequations}
\label{eq:opt9}
\begin{align}
\underset{\underline{\bm\theta}}{\max} \quad & f_{10}(\underline{\bm\theta}) = -\underline{\bm\theta}^{\sf H} \mathbf A \underline{\bm\theta} + 2\Re\{\underline{\bm\theta}^{\sf H} \mathbf b\} \\
\text{s.t.} \quad &  \vert \underline{\theta}_{\kappa} \vert^2 \leq 1, \quad \forall \kappa=1,2,\cdots, M_{\text{tot}}, \label{eq:condition_cvx}
\end{align}
\end{subequations}
\noindent where $M_{\text{tot}} = MG$. Problem (\ref{eq:opt9}) is a convex optimization problem. Accordingly, we can resort to the Lagrangian dual decomposition method to solve (\ref{eq:opt9}), since the strong duality holds. The Lagrange dual of (\ref{eq:opt9}) can be written as
\begin{subequations}
\label{eq:opt10}
\begin{align}
\underset{ \bm\zeta }{\min} \quad & f_{\mathcal D}(\bm\zeta) = \underset{\underline{\bm\theta}}{\sup} \quad f_{\mathcal L}(\underline{\bm\theta}, \bm\zeta) \label{eq:LDD} \\
\text{s.t.} \quad &  \zeta_{\kappa} \geq 0, \quad \forall \kappa=1,2,\cdots, M_{\text{tot}}. \label{eq:condition_cvx2}
\end{align}
\end{subequations}
{\noindent}In (\ref{eq:LDD}),
\begin{align}
f_{\mathcal L}(\underline{\bm\theta}, \bm\zeta) = f_{10}(\underline{\bm\theta}) - \sum_{\kappa = 1}^{M_{\text{tot}}} \zeta_{\kappa} \left(\underline{\bm\theta}^{\sf H} \mathbf e_{\kappa} \mathbf e_{\kappa}^{\sf H} \underline{\bm\theta} - 1 \right),
\label{eq:Lagrangian}
\end{align}
where $\mathbf e_{\kappa} \in \mathbb R^{M_{\text{tot}} \times 1}$ is the elementary vector with one in the $\kappa$-th position and zeros elsewhere; $\bm\zeta = \left[\zeta_1, \zeta_2, \cdots, \zeta_{M_{\text{tot}}} \right]^{\sf T}$ collects the dual variables associated with constraint (\ref{eq:condition_cvx}).

By setting ${\partial f_{\mathcal L}}/{\partial \underline{\bm\theta}}=0$, the optimal $\underline{\bm\theta}$ can be obtained as
\begin{align}
\hat{\underline{\bm\theta}} &= \Big(\mathbf A + \sum_{\kappa = 1}^{M_{\text{tot}}} \zeta_{\kappa} \mathbf e_{\kappa} \mathbf e_{\kappa}^{\sf H} \Big)^{-1} \cdot \mathbf b = \mathbf D(\bm\zeta) \mathbf b, \label{eq:opt_theta}
\end{align}
where $\mathbf D(\bm\zeta) = \left( \mathbf A + \text{diag}\left(\bm\zeta\right) \right)^{-1}$. The optimal value of ${\bm\zeta}$ is given in the following lemma.

\begin{lemm}
The optimal value of ${\bm\zeta}$ in (\ref{eq:opt_theta}) can be determined by
\begin{align}
\hat{\bm\zeta} = \left\{\zeta_{\kappa} \geq 0: \Big[\mathbf D(\bm\zeta) \mathbf b \Big] \odot \Big[\mathbf D(\bm\zeta) \mathbf b \Big]^{\ast} = \mathbf 1 \right\}, \label{eq:optimal_zeta}
\end{align}
which indicates that the solution attained by the constraint relaxation-based method can guarantee that  the unit-modulus constraints $\underline{\bm\theta}^{\sf H} \mathbf e_{\kappa} \mathbf e_{\kappa}^{\sf H} \underline{\bm\theta} = 1$ are satisfied.
\end{lemm}
\begin{IEEEproof}
See Appendix \ref{prf3}.
\end{IEEEproof}

Substituting (\ref{eq:opt_theta}) into (\ref{eq:opt10}) yields
\begin{align}
f_{\mathcal D}(\bm\zeta) &= \mathbf b^{\sf H} \mathbf D(\bm\zeta) \mathbf b + \text{tr}(\text{diag}\left(\bm\zeta\right)). \label{eq:dual}
\end{align}
In other words, (\ref{eq:opt10}) can be recast as a semi-definite programming (SDP) problem using the Schur complement:
\begin{subequations}
\label{eq:opt11}
\begin{align}
\underset{ \bm\zeta,  \epsilon}{\max} \quad & \epsilon-\text{tr}(\text{diag}\left(\bm\zeta\right)) \\
\text{s.t.} \quad &   \begin{bmatrix}
\mathbf A + \text{diag}\left(\bm\zeta\right) & \mathbf b \\
\mathbf b^{\sf H} & -\epsilon
\end{bmatrix} \succcurlyeq \mathbf 0,
\end{align}
\end{subequations}
which can be solved efficiently by using standard CVX toolboxes.

\begin{algorithm}[htbp]
\small
\caption{The proposed alternating iterative framework.}
\label{alg:A1}
\begin{algorithmic}[1]
\REQUIRE {Set feasible values of $\{\mathbf P^{(0)}, \mathbf \Phi_g^{(0)}\}$ and $t=0$.}
\REPEAT
\STATE {Set $t \leftarrow t+1$;}
\STATE {Update $\alpha_k^{(t)} = \gamma_k$ by (\ref{eq:sinr});}
\STATE {Update $\beta_k^{(t)}$ and $\mathbf p_k^{(t)}$ by (\ref{eq:optimal_beta}) and (\ref{eq:optimal_p}), respectively;}
\STATE {Construct ${\mathbf \Theta}^{(t)}$ and ${\mathbf V}_{k,j}^{(t)}$ by (\ref{eq:v21}) and (\ref{eq:v22});}
\STATE {Update $\rho_k^{(t)}$ by (\ref{eq:rho}) to compute $\mathbf A^{(t)}$ and $\mathbf b^{(t)}$;}
\STATE {Update $\bm\zeta^{(t)}$ by solving problem (\ref{eq:opt11}) to obtain $\underline{\bm\theta}^{(t)}$ by (\ref{eq:opt_theta});}
\STATE {With given $\underline{\bm\theta}^{(t)}$, update $\mathbf \Phi_g^{(t)}$.}
\UNTIL {The function (\ref{eq:sumR2}) converges.}
\STATE {Perform the quantized phase projection.}
\end{algorithmic}
\end{algorithm}

Furthermore, we can leverage the quantized phase projection method \cite{8982186} to discretize the solutions.
The proposed alternating iterative framework to solve Problem (\ref{eq:sumR2}) is provided in Algorithm \ref{alg:A1}.
To illustrate the convergence of the alternating iterative approach, we provide the following proposition.

\begin{prop}
The problem in (\ref{eq:sumR2}) converges when the alternating iterative algorithm is used.
\end{prop}

\begin{IEEEproof}
See Appendix \ref{prf4}.
\end{IEEEproof}

\section{Numerical Results}\label{section:sim}
Without loss of generality, the weights $\omega_k$ are set to be equal in all the simulations. According to \cite{wang2019intelligent} and \cite{6834753}, the channel gain is taken as $\nu_k \sim \mathcal{CN}(0, 10^{-0.1 \mathrm{PL}(r)})$ where $\mathrm{PL}(r)=\varrho_a + 10\varrho_b \lg(r)+ \xi$ with $\xi \sim \mathcal{N}(0,\sigma_{\xi}^2)$\footnote{The distance between transmitter and receiver is denoted as $r$, which can be calculated according to the simulated system layout.}. The channel realizations are produced by setting $\sigma_{\text u}^2=-85$ dBm, $\varrho_{\text t}=9.82$ dBi, $\varrho_{\text r}=0$ dBi, $\varrho_{\text a}=61.4$, $\varrho_{\text b}=2$, and $\sigma_{\xi}=5.8$ dB.
The BS with $N=32$ antennas is located at the origin and the MUs are uniformly and randomly distributed in an circle centered at $(40\ \mathrm m, 0\ \mathrm m)$ with radius $10\ \mathrm m$. Here, we consider $G=2$ IRS units located at $(40\ \mathrm m, 30\ \mathrm m)$ and $(30\ \mathrm m, 40\ \mathrm m)$. Unless otherwise stated, the total transmit power is set to $P_{\max}=30$ dBm \cite{wang2019intelligent}.

\begin{figure}[t]
    \centering{}\includegraphics[scale=0.48]{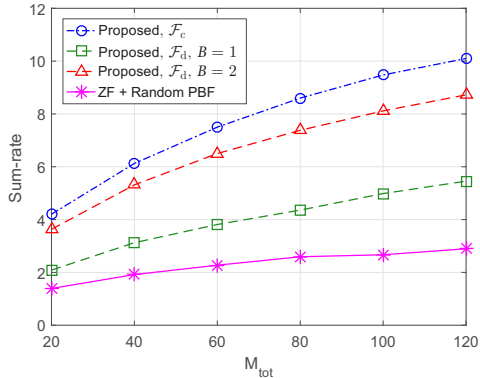}
    \caption{Achievable sum-rate versus IRS size with $N_{\rm p}=2$ and $K=2$.}
    \label{fig:sim1}
\end{figure}

\begin{figure}[t]
	\centering{}\includegraphics[scale=0.48]{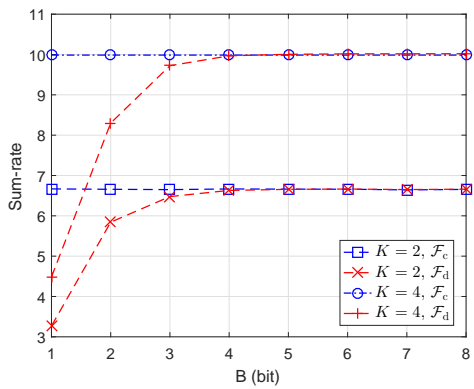}
	\caption{Achievable sum-rate versus $B$ with $M_{\text{tot}}=20$ and $N_{\rm p}=2$.}
	\label{fig:sim2}
\end{figure}

Fig. \ref{fig:sim1} shows the sum-rate versus the number of reflecting elements where $M_{\rm az}=10$ and $M_{\rm el}$ increases from 1 to 6. Zero-forcing (ZF) precoding with a random PBF method is simulated as the baseline. Clearly, the proposed algorithm outperforms the baseline method in terms of sum-rate. The sum-rate increases, as $M_{\text{tot}}$ increases from 20 to 120. Moreover, the proposed algorithm with 1-bit phase shifters at the IRS can consistently outperform the baseline method, and the gain becomes significant when 2-bit phase shifters are adopted.

\begin{figure}[t]
	\centering{}\includegraphics[scale=0.48]{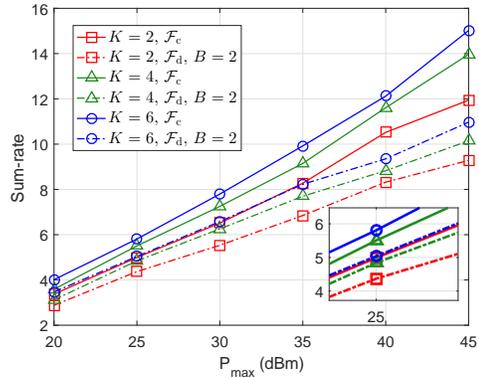}
	\caption{Achievable sum-rate versus $P_{\max}$ with $M_{\text{tot}}=20$ and $N_{\rm p}=2$.}
	\label{fig:sim3}
\end{figure}

\begin{figure}[t]
	\centering{}\includegraphics[scale=0.48]{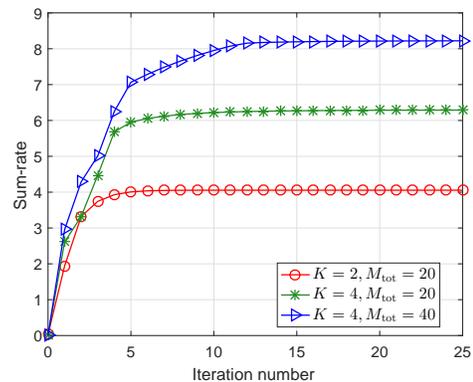}
	\caption{Convergence behavior of the proposed algorithm.}
	\label{fig:sim4}
\end{figure}

To investigate the phase resolution of IRS required to archive acceptable system performance, Fig. \ref{fig:sim2} shows the performance gap between phase-shifter of infinite resolution and that of various quantization bits. Each IRS unit size is set as $M_{\rm az}=10$ and $M_{\rm el}=2$. It is observed that the sum-rate performance gap gradually decreases with $B$ increasing from 1 bit to 8 bits. When $B=3$, the performance gap can reach 0.17 and 0.26 for 2-user case and 4-user case, respectively. We conclude that only 4-bit phase shifter can achieve the equivalent performance gain as continuous phase.

Fig. \ref{fig:sim3} assesses the impact of the transmit power on the sum-rate of our proposed algorithm. We set $M_{\rm az}=10$ and $M_{\rm el}=2$. The red, green and blue lines indicate the 2-user, 4-user, and 6-user scenarios, respectively. As depicted in the figure, with the transmit power increasing from 20 dBm to 45 dBm, all of the multi-user cases exhibit the same upward trend. As the number of users grows in the network, the sum-rate attained under the infinite resolution elements and the same transmit power also increases. The tendency of 2-bit resolution case is consistent with that of the infinite resolution case.

Fig. \ref{fig:sim4} shows the convergence trend of the proposed alternating iterative algorithm. The convergence of our proposed algorithm is confirmed in multiple simulated cases. When $K=2$ and $M_{\text{tot}}=20$, our proposed method converges after about 5 iterations. As the number of users or the value of $M_{\text{tot}}$ increases, we can see the convergent sum-rate of the proposed algorithm grows.
To conclude, the proposed algorithm can achieve fast convergence.

\section{Conclusion}\label{section:con}
To support multi-user transmission in the mmWave blind spots, a distributed deployment of IRS units was introduced to create the high-rank mmWave channels. In this paper, an alternating iterative algorithm was designed to maximize the weighted sum-rate of the IRS-aided mmWave systems. The algorithm jointly optimizes ABF and PBF with closed-form solutions at each iteration. Simulation results corroborated the feasibility and effectiveness of the proposed scheme.

\begin{appendices}
\section{Proof of Proposition 1}\label{prf1}
Note that $f_2$ is a concave differentiable function with respect to $\alpha_k$ when given $\gamma_k$. Thus, setting $\frac{\partial f_2}{\partial \alpha_k}$ to zero yields $\hat{\alpha}_k=\gamma_k$. Based on this fundamental result, substituting $\hat{\bm\alpha}$ back in the objective function of (\ref{eq:fopt}) can be recast to (\ref{eq:sumR2}). As such, the optimal objective values of these two problems are equal and their equivalence is established. \QEDA

\section{Proof of Lemma 1}\label{prf2}
For notational brevity, rearranging (\ref{eq:optimal_p}) leads to
\begin{align}
\hat{\mathbf P} = \left(\mu \mathbf I_N + \mathbf X \right)^{-1} \mathbf Y \mathbf \Lambda,
\label{eq:pp}
\end{align}
where $\mathbf X = \sum_{i=1}^{K} \vert \beta_i \vert^2 \tilde{\mathbf h}_i \tilde{\mathbf h}_i^{\sf H}$, $\mathbf Y = [\tilde{\mathbf h}_1, \tilde{\mathbf h}_2, \cdots, \tilde{\mathbf h}_K ]$, and $\mathbf \Lambda = \textup{diag}\left(\left[\sqrt{\bar{\alpha}_1} {\beta}_1, \cdots, \sqrt{\bar{\alpha}_K} {\beta}_K \right]\right)$.
Then, the derivation of transmit power with respect to $\mu$ is
\begin{align}
\frac{\partial \textup{tr}\left(\mathbf{P} \mathbf{P}^{\sf H}\right)}{\partial \mu} &= \textup{tr} \left(
\bigg(\frac{\partial \textup{tr} \left(\mathbf{P} \mathbf{P}^{\sf H}\right)}{\partial \mathbf P}\bigg)^{\sf H} \frac{\partial \mathbf P}{\partial \mu}
\right) \nonumber \\
&= -2 \cdot \textup{tr} \left(
\mathbf{P}^{\sf H} \left(\mu \mathbf I_N + \mathbf X\right)^{-1} \left(\mu \mathbf I_N + \mathbf X\right)^{-1} \mathbf Y \mathbf \Lambda
\right) \nonumber \\
&= -2 \cdot \textup{tr} \left(
\mathbf{P}^{\sf H} \left(\mu \mathbf I_N + \mathbf X\right)^{-1} \mathbf{P}
\right).
\end{align}
which shows that (\ref{eq:pp}) is monotonically decreasing as $\mu$ increases. \QEDA

\section{Proof of Lemma 2}\label{prf3}
Using the chain rule in matrix differentiation, we have
\begin{align}
\frac{\partial f_{\mathcal D} }{\partial \zeta_{\kappa}} &= 1-\text{tr}\left[ \mathbf D(\bm\zeta) \mathbf b \mathbf b^{\sf H} \mathbf D(\bm\zeta)  \frac{\partial \left( \mathbf A + \text{diag}\left(\bm\zeta\right) \right)}{\partial \zeta_{\kappa}} \right]  \nonumber \\
&= 1- \left[ \mathbf D(\bm\zeta) \mathbf b \mathbf b^{\sf H} \mathbf D(\bm\zeta) \right]_{\kappa, \kappa}. \label{eq:dual_derivative}
\end{align}
Setting (\ref{eq:dual_derivative}) to zero yields
\begin{align}
[\mathbf D(\bm\zeta) \mathbf b ] \odot [\mathbf D(\bm\zeta) \mathbf b ]^{\ast} = \underline{\bm\theta} \odot \underline{\bm\theta}^{\sf H} = \mathbf 1.
\end{align}
This completes our proof. \QEDA

\section{Proof of Proposition 2}\label{prf4}
Since the optimum solution can be attained at each iteration, we have
\begin{align}
f_1(\mathbf P^{(t+1)}, \mathbf \Phi_g^{(t+1)}) &= f_2(\mathbf P^{(t+1)}, \mathbf \Phi_g^{(t+1)}, \bm \alpha^{(t+1)}) \nonumber \\
&\ge f_2(\mathbf P^{(t+1)}, \mathbf \Phi_g^{(t+1)}, \bm \alpha^{(t)}) \nonumber \\
&\ge f_2(\mathbf P^{(t)}, \mathbf \Phi_g^{(t+1)}, \bm \alpha^{(t)}) \nonumber \\
&\ge f_2(\mathbf P^{(t)}, \mathbf \Phi_g^{(t)}, \bm \alpha^{(t)}) \nonumber \\
&= f_1(\mathbf P^{(t)}, \mathbf \Phi_g^{(t)}).
\end{align}
where $t$ is the iteration index in Algorithm \ref{alg:A1}.
Since the WSM problem is bounded above due to the power constraints \cite{8314727}, the original objective function is monotonically nondecreasing after each iteration. \QEDA

\end{appendices}

\bibliographystyle{IEEEtran}

\begin{thebibliography}{10}
\providecommand{\url}[1]{#1}
\csname url@samestyle\endcsname
\providecommand{\newblock}{\relax}
\providecommand{\bibinfo}[2]{#2}
\providecommand{\BIBentrySTDinterwordspacing}{\spaceskip=0pt\relax}
\providecommand{\BIBentryALTinterwordstretchfactor}{4}
\providecommand{\BIBentryALTinterwordspacing}{\spaceskip=\fontdimen2\font plus
\BIBentryALTinterwordstretchfactor\fontdimen3\font minus
  \fontdimen4\font\relax}
\providecommand{\BIBforeignlanguage}[2]{{%
\expandafter\ifx\csname l@#1\endcsname\relax
\typeout{** WARNING: IEEEtran.bst: No hyphenation pattern has been}%
\typeout{** loaded for the language `#1'. Using the pattern for}%
\typeout{** the default language instead.}%
\else
\language=\csname l@#1\endcsname
\fi
#2}}
\providecommand{\BIBdecl}{\relax}
\BIBdecl

\bibitem{8811733}
Q.~{Wu} and R.~{Zhang}, ``Intelligent reflecting surface enhanced wireless
  network via joint active and passive beamforming,'' \emph{IEEE Trans.
  Wireless Commun.}, pp. 1--1, Aug. 2019.

\bibitem{6847111}
A.~{Alkhateeb}, O.~{El Ayach}, G.~{Leus}, and R.~W. {Heath}, ``Channel
  estimation and hybrid precoding for millimeter wave cellular systems,''
  \emph{IEEE J. Sel. Topics Signal Process.}, vol.~8, no.~5, pp. 831--846, Oct.
  2014.

\bibitem{7306533}
Z.~{Gao}, L.~{Dai}, D.~{Mi}, Z.~{Wang}, M.~A. {Imran} \emph{et~al.}, ``{MmWave}
  massive-{MIMO}-based wireless backhaul for the {5G} ultra-dense network,''
  \emph{IEEE Wireless Commun.}, vol.~22, no.~5, pp. 13--21, Oct. 2015.

\bibitem{8741198}
C.~{Huang}, A.~{Zappone}, G.~C. {Alexandropoulos}, M.~{Debbah}, and C.~{Yuen},
  ``Reconfigurable intelligent surfaces for energy efficiency in wireless
  communication,'' \emph{IEEE Trans. Wireless Commun.}, vol.~18, no.~8, pp.
  4157--4170, Aug. 2019.

\bibitem{8683145}
Q.~{Wu} and R.~{Zhang}, ``Beamforming optimization for intelligent reflecting
  surface with discrete phase shifts,'' in \emph{Proc. IEEE Int. Conf. Acoust.,
  Speech Signal Process. (ICASSP)}, Brighton, United Kingdom, May 2019, pp.
  7830--7833.

\bibitem{8723525}
M.~{Cui}, G.~{Zhang}, and R.~{Zhang}, ``Secure wireless communication via
  intelligent reflecting surface,'' \emph{IEEE Wireless Commun. Lett.}, pp.
  1--1, May 2019.

\bibitem{8982186}
H.~{Guo}, Y.~{Liang}, J.~{Chen}, and E.~G. {Larsson}, ``Weighted sum-rate
  maximization for reconfigurable intelligent surface aided wireless
  networks,'' \emph{IEEE Trans. Wireless Commun.}, pp. 1--1, 2020.

\bibitem{8746155}
Y.~{Han}, W.~{Tang}, S.~{Jin}, C.~{Wen}, and X.~{Ma}, ``Large intelligent
  surface-assisted wireless communication exploiting statistical {CSI},''
  \emph{IEEE Trans. Veh. Technol.}, vol.~68, no.~8, pp. 8238--8242, Aug. 2019.

\bibitem{wang2019intelligent}
P.~Wang, J.~Fang, X.~Yuan, Z.~Chen, H.~Duan \emph{et~al.}, ``Intelligent
  reflecting surface-assisted millimeter wave communications: Joint active and
  passive precoding design,'' \emph{arXiv preprint arXiv:1908.10734}, 2019.

\bibitem{6717211}
O.~E. {Ayach}, S.~{Rajagopal}, S.~{Abu-Surra}, Z.~{Pi}, and R.~W. {Heath},
  ``Spatially sparse precoding in millimeter wave {MIMO} systems,'' \emph{IEEE
  Trans. Wireless Commun.}, vol.~13, no.~3, pp. 1499--1513, Mar. 2014.

\bibitem{6834753}
M.~R. {Akdeniz}, Y.~{Liu}, M.~K. {Samimi}, S.~{Sun}, S.~{Rangan} \emph{et~al.},
  ``Millimeter wave channel modeling and cellular capacity evaluation,''
  \emph{IEEE J. Sel. Areas Commun.}, vol.~32, no.~6, pp. 1164--1179, Jun. 2014.

\bibitem{8580675}
S.~{Hu}, K.~{Chitti}, F.~{Rusek}, and O.~{Edfors}, ``User assignment with
  distributed large intelligent surface ({LIS}) systems,'' in \emph{Proc. IEEE
  Annu. Int. Symp. Pers., Indoor, Mobile Radio Commun. (PIMRC)}, Bologna,
  Italy, Sep. 2018, pp. 1--6.

\bibitem{8314727}
K.~{Shen} and W.~{Yu}, ``Fractional programming for communication systems-part
  {I}: Power control and beamforming,'' \emph{IEEE Trans. Signal Process.},
  vol.~66, no.~10, pp. 2616--2630, May 2018.

\end{thebibliography}

\end{document}